\documentclass[12pt]{article}

\usepackage{graphicx}
\usepackage[latin1]{inputenc}
\usepackage{amsfonts}
\usepackage{amsmath}

\begin{document}

	\title{\vspace{-2.5cm}{\Large \bf 
			Embracing undecidability: Cognitive needs and theory evaluation}
	}
	\author{{\bf Andr\'e C. R. Martins}
	}
	
	\date{}
	
	\maketitle

\begin{abstract}
    There are many ways we can not know. Even in systems that we created ourselves, as, for example, systems in mathematical logic, Go\"edel and Tarski's theorems impose limits on what we can know. As we try to speak of the real world, things get even harder. We want to compare the results of our mathematical theories to observations, and that means the use of inductive methods. While we can demonstrate how an ideal probabilistic induction should work, the requirements of such a method include a few infinities. Furthermore, it would not be even enough to be able to compute those methods and obtain predictions. There are cases where underdeterminacy might be unavoidable, such as the interpretation of quantum mechanics or the current status of string theory. Despite that, scientists still behave as if they were able to know the truth. As it becomes clear that such behavior can cause severe cognitive mistakes,  the need to accept our limits, both our natural human limits and the limits of the tools we have created, become apparent. This essay will discuss how we must accept that knowledge is almost only limited to formal systems. Moreover, even in those, there will always be undecidable propositions. We will also see how those questions influence the evaluation of current theories in physics.
    
\end{abstract}

\section{Introduction}
\label{intro}

    The attempts, at the start of the XX$^{th}$ century, to base mathematical knowledge (and from there, all human knowledge) on logic was a brave effort. The attempts might have failed, but they were an honest and very much needed effort to diminish the influence of human subjectivity in our conclusions. However, instead of obtaining reliable recipes on how we can prove and know statements, we have learned there are limits to what we can achieve. Those happen even in mathematical logic. In pure mathematical systems, we decide the fundamental axioms. There is no need to match the real world. While those systems are artificial worlds we create, that does not mean we can know every statement in those worlds. Assuming we did not introduce any inconsistencies in our axioms, we can prove some statements to be true and others to be false. However, since Go\"edel\cite{godel62a} and Tarski\cite{tarski83a}, we know there will always be other statements we can not decide. 
    
    In principle, that could lead to no problems when we try to describe the real world, except for a radical program of logicism. Indeed, we assume that our theories can not decide all the details of the universe. We usually assume theories also need boundary and initial conditions. Gravitation theories do not tell us how many planets exist in a solar system; theories in physics are supposed to have undecidable components that we will observe and not deduct. Auxiliary hypothesis \cite{duhem89a,quine53a}are unavoidable. They are not considered a problem in that context. Indeed, we often use them to rescue theories when experiments and theories do not agree. Take the case of both dark matter and dark energy. We do not know what they are, nor if they exist. However, they are needed as additional terms we add to the equations of movement of stars in a galaxy and how fast galaxies move from us as a function of their distances. Interpretations of those terms exist. One of them might be right. However, we do not know. The error might be in our theories just as it can also be in our observations.
    
    As we try to learn about the real world, things get more complicated. New levels of impossibility appear. At least, our theories can make predictions, even if they leave statements as undecidable. However, we have to compare those predictions against what we observe, and predictions are never exact, even for deterministic theories. Experiments have errors. The constants in any theory are only known up to limited precision. That means that even our most reliable predictions are probabilistic. In chaotic systems, it is even common to have situations where our best predictions do not limit the possible states of the system to a narrower range than what we knew before making the prediction. Some regularities might arise and, when that is the case, we can check them. Nevertheless, comparing observations to a distribution that is too wide might not provide as much information as we would have liked.
    
    Worse still, the predictions of a theory are not a well-defined set. As in the solar system example, we can only try to apply any gravitational theory, Newtonian, general relativity, or others, if we have extra information. In that case, we need as the auxiliary hypotheses the characteristics, position, and momenta of each mass in the system. A proper and complete prediction must also include the process of data gathering. Possible errors in the experimental setup must be considered. Anything that might interfere with the results -- or their precision -- will affect the final probability distributions. At least, that is the ideal and correct case we get from probabilistic methods. However, considering all possibilities means including infinite alternatives, making the problem non-computable. 
    
    The amount of difficulties we find when we try to create logically sound ways to estimate theories is so large that we considering alternatives might be necessary sometimes. However, without the guidance of logically sound methods, we might be left with our natural reasoning. It is, therefore, worth discussing what we have learned from the most recent experiments in our cognitive abilities.

	\section{Cognition and Beliefs}
	
	We have ideas we like. Furthermore, we too often define ourselves in terms of the ideas we defend. If we are talking about preferences, those identity-defining ideas are perfectly fine. They are, indeed, one of the bases of a democratic debate. Nevertheless, we also pick ideas we prefer about how the world is. However, that is not a problem where we can choose. Whatever our preferences might be, the universe will still be the way it is. Our opinions, here, do not matter.
	
	Since we can learn and change our minds, that tendency to pick preferred ideas about the real world might not have been a problem. If it were just a quirk, with no damaging consequences, we could ignore it. Unfortunately, that is not the case.
	
	Cognitive experiments have been consistently showing we are not born logicians. Instead, when using our natural skills, we perform poorly even in simple logical and probabilistic problems \cite{baron}. While those errors are quite common, they are not random and do not mean we are incompetent. Indeed, our mistakes can be understood and classified according to their causes. One of those causes is the use of fast heuristics \cite{gigerenzerbrighton09a}. Heuristics are simple rules of thumbs that fall much shorter of a logical analysis of the problem. They are, however, much faster. In everyday life, there are often situations where the speed of decision might matter far more than accuracy. 
	
	A second reason for how we do not behave in experiments as well as we should is related to the experiments themselves. Experimental setups are usually simplified versions of actual problems. There is evidence our minds work in ways that are similar to a Bayesian inference \cite{tenenbaumetal07}. In the experiments, scientists provide information to their subjects, and they expect those subjects would treat that information as entirely reliable. However, there is no utterly reliable information in the real world. If one uses the information as data and introduces the possibility of errors and deceptions, a Bayesian analysis of several experiments show our behavior is not as wrong as it might have initially looked \cite{martins06}.
	
	That does not mean, however, all our mistakes are functional adaptations to finding the best answers in the real world. We are overconfident about our reasoning skills~\cite{oskamp65a}. Not only that but, when we get more information, we can get more confident even when we do not become better at our estimates \cite{halletal07a}. That seems to serve no purpose if what we care about is finding correct answers. However, being confident helps us convince others \cite{tiedenslinton01a}. That is a crucial point.
	
	It seems our argumentation skills did not evolve to help us arrive at the best answers. Instead, they are probably efficient adaptations for fitting in our social groups \cite{merciersperber11a,Mercier2017}. We evolved our abilities so that we could convince others and get into positions of power. Failing that, we conform and accept the views of our groups. Fitting in our social group can be much more important than finding the best idea, at least when the ideas of our group are not an actual disaster. Instead, we use our reasoning to defend our identity-defining ideas \cite{kahan13a,Kahan2017b}.
	Furthermore, being intelligent seems not to help to protect anyone from that problem. Quite the opposite. It seems that better numeracy skills might, instead, help each person defend her preferences better \cite{KAHAN2017}. That can make competent people more polarized than those without the same skills even when data does not support such polarization.
	
	That suggests that having strong beliefs can be very detrimental to our natural reasoning skills. At least if we mean to use them to reach correct conclusions \cite{martins16a}. Scientists do get attached to their ideas. Indeed, debates in science can be as heated as in any other activity. So, it is no surprise we should fall prey to the same biases when debating scientific issues. That strongly suggests we should not trust our natural reasoning skills whenever that is possible. Furthermore, we should limit ourselves to the best logical and mathematical methods we have. As such, even when we reach problems that require infinite capabilities, a better understanding of what logic and inductive methods suggest can help us learn where we might be wrong in our current practices.

	\section{Induction and Bayesian methods}

	Evaluating theories can not be done based on deductions alone. A mathematical theory tells what we should observe if it is right. If our observations agree with the theory, that does not mean the theory is correct. It just means it is compatible with the observations. That is not all, however. If the observations and theory disagree, that also does not prove the theory wrong. Something is wrong, of course, but not necessarily the theory. The error might be in the experiment, in the auxiliary hypothesis, even in the theoretical demonstrations. Physicists might like the Popperian notion that we can show theories can to be wrong \cite{popper59}. While that idea can be an excellent approximation in some cases \cite{Martins2017}, it is not correct.
	
	Comparing observation and predictions can only be done, given our current knowledge, using inductive methods. Interestingly, there are known ways to treat the problem of the plausibility of a statement. We can work from reasonable and straightforward desiderata and prove, up to a trivial transformation, that, if we want to follow them, we must use probabilities. That also means we must update our plausibilities using Bayes theorem \cite{cox61a}. Other demonstrations that Bayesian methods are the correct way to update plausibilities include the weak ``Dutch book'' argument and also maximum entropy methods \cite{catichagiffin07a}.
	
	Bayesian methods, however, are not entirely accepted \cite{kaplan02}. While logically sound, they suffer from serious uncomputability problems. Indeed, while it is easy to find books on Bayesian methods, they never propose using the method to its full extent. Approximations are always presented as valid methods. They are indeed good ways to obtain educated guesses. However, when one makes an approximation to a logically correct theory, inconsistencies can appear. They do appear. Many criticisms and apparent problems we find in the literature are a consequence of not using the complete -- albeit impossible -- method \cite{Martins2020inpress}. That, of course, does not mean Bayesian methods are wrong. It only means they might not be applicable in practice in every problem. That is the case, and finding ways to obtain better approximations is an open research question.
	
	One of the evident problems in applying Bayesian methods is to determine our initial probabilistic opinion on the problem, the prior distribution. As probability laws do not determine those values, people can see them as open. They do depend on what each person knows. That can be very different from one individual to another. Therefore, while there are attempts at calculating objective priors, they are often considered subjective. That might sound as meaning that each person can choose them freely. However, that is only true before one learns anything. In principle, a person should pick a prior when she starts learning about the world. From there, she must update that initial guess with everything she has learned in her entire life. While not precisely infinite information, such an estimate is not computable in practical terms. Hence, even competent statisticians end up using guesses. However, it is crucial to recognize that those methods are not perfect implementations.
	
	Another fundamental uncomputable aspect of Bayesian methods comes from the need to include every possible theory and their variations, including those nobody has ever thought \cite{fitelsonthomason08}. That is needed if one wants to obtain actual probability values. Comparing two complete sets of ideas can still be done if we ask how the odds-ratio between the two sets evolve. In that case, renormalization constants cancel out, and the result does not depend on ideas outside the set. While useful, this property does not solve the problem of evaluating theories. As we will see, it allows us to make comparisons, but those still depend on assumptions that are not part of the compared theories.

	\section{Theories in physics}

	One central cause that makes even comparing two theories using Bayesian methods often impossible is the fact that theories alone do not make probabilistic predictions. If we keep a theory unchanged, but we alter our assumptions about details like experimental errors or initial conditions, predictions change. In principle, the prediction of a theory should include distributions for every possibility associated with errors, possible initial and boundary conditions, details on other theories and hypotheses, and everything else that might influence the outcome. Each possibility should be considered and introduced with the proper weight, pre-obtained from earlier inferential work. 
	
	Of course, we can not know all theories and make predictions from them all. That means it is not possible to calculate the correct predictions of any theory. We can, as always, work with the assumptions we already know and hope whatever we disregard is very improbable. If we are lucky, we might get reliable approximations. However, we have no way to estimate the consequences of disregarding ideas we do not even know.
	
	When comparing specific sets of assumptions, however, Bayesian methods can provide remarkably precise answers. With a little care, we can even investigate if changing a few hypotheses would cause essential changes. Take, for example, the evidence in favor of general relativity when compared to Newtonian mechanics. If we look at the historical numbers that suggested general relativity was a better theory, the difference is colossal \cite{clemence47a}. Indeed, take only the case of the precession of the perihelion of Mercury. Given the average observations, the predictions of both theories, and the standard deviations associated with those numbers, if we assume errors follow a Normal distribution, the evidence in favor of general relativity amounts to an odds-ratio of staggering $10^{445}$. If one starts thinking there is only one in a million chance general relativity is better, that ratio will turn that initial believe into a one in $10^{439}$ chances in favor of general relativity.
	
	That calculation, of course, makes several common assumptions, such as the known structure of the solar system, with its planets and known orbits, for example. It also assumes the standard versions of both Newtonian mechanics and general relativity. Any corrections to those theories are left out. There is, however, one hypothesis we can quickly check for its effect. Those numbers assume a Normal distribution for the experimental errors. Normal distributions, however, are very thin-tailed functions. When there are several sources of observational errors, they tend to provide a first approximation. More than that, they work much better at central values and start failing first at their thin tails. 
	
	If we investigate what happens if we replace the distribution with another with much fatter tails, the difference is also surprising. Assuming a $t$-distribution with only 10 degrees of freedom, the support for general relativity falls from $10^{445}$ to $10^{12}$. It is still a significant change in opinion, but the awe-inspiring support from only observing Mercury evaporates. Knowing which distribution to use, however, would require a complete study of how all values were calculated. That would mean estimates on the possible source of errors both in the experiment and in the theory. We would also need to calculate how those errors propagate. The most likely scenario, likely from nothing more than an educated guess, is that errors should have a distribution with tails fatter than a Normal, but not necessarily so fat. That would put the odds evidence in favor of general relativity between those two numbers.
	
	Of course, that is the evidence from one case alone. General relativity has been subject to a variety of other tests and, so far, has shown to be superior in every one of them. That means the support in its favor today is incredibly strong (if we assume standard auxiliary hypothesis). The support in favor of our best physical theories might vary. However, when predictions are far enough when compared to the experimental errors, we can feel entirely justified at saying that one theory is better. Induction will tell us there are chances we might be wrong, but chances like one in $10^{100}$ are identical to zero to anyone who is not a mathematician. The difference is purely technical, with no consequences in the real world. 
	
	It is easy to see why physicists tend to adopt realism as their favorite philosophical posture. The precision in our measurements and how they match some observations have only improved since Wigner \cite{wigner60a} labeled mathematical effectiveness as unreasonable. However, what those calculations tell us is not that our theories are correct or that mathematics rules the universe. Those are possibilities, but they belong to pure metaphysics. Nothing in our experiments tell that. What we do know is that our theories describe what we have observed incredibly well. However, even the term incredibly only makes sense when compared to natural human abilities. No matter how many digits we might get right for some theories, they will be nothing compared to the infinite number of possible digits. We should also not forget that we were only able to obtain that strength of evidence by making extra assumptions. That means excluding other possibilities, as we are incapable of theoretical omniscience. Calculating all likelihoods and theories is impossible. We do not know how well whatever we have left out might have performed.
	
	On the other hand, there are other fields of scientific knowledge where getting probabilistic predictions from theories, even making many assumptions, is an almost impossible problem. While it is quite clear what we should assume as the characteristics of the solar system, assuming how each part influences each other in a social problem is no simple task. There are just too many possible variables. Furthermore, there too many theories on how we behave, as individuals and in society. None of those theories work so well that we can discard the others. As theories and hypotheses become too many, predicting outcomes is no longer possible. In that case, it is reasonable that many competing ideas can survive even as we observe how social systems work in the world. The need for theoretical work is clear since we can not trust our natural reasoning. However, while mathematical and even computational models are more reliable than human cognition, the impossibility of making precise predictions means we should remain in doubt about which theories can describe those problems correctly. That suggests that, just as it was natural for physicists to embrace realism, for researchers in social sciences, the "everything goes" from post-modernism might seem just as appealing. Both positions have no logical justification, but they do describe the daily life of typical research in those areas well.
	
	If we leave historical analysis behind and move to open problems in physics, things are no longer as easy as they seemed to be. In quite a few open questions, physics now suffers from a problem of undecidability. Take quantum mechanics. The theory works incredibly well. It is one of the best cases of success in physics. However, if we try to understand why quantum mechanics works the way it does, we have to check its competing interpretations. Those interpretations, except for the question of the impossibility of local hidden variables, are not experimentally verifiable. The interpretations provide the same equations, the same predictions. That means that whatever we observe, data will not change our initial opinions. At all. The interpretations are underdetermined, exactly undecidable.
	
	The same happens with string theory. Despite attempts to find ways to update probabilities with no data support \cite{dawid13a}, those proposals are only suggestions for picking priors and heuristical advice, based on past cases, on which path a scientist should take to tackle a new problem \cite{Martins2017}. However, the existence of more than one theory that provides the same predictions is not a practical problem. Indeed, when estimating how a system will behave, the probabilities of each competing theory with identical predictions will add up, making their outcome more likely. A similar argument exists using alternative methods from Solomonoff induction \cite{Martins2020inpress}.
	
	Understanding our limitations is crucial. Physicists have been shielded from those problems while their theories provided very distinct and decidable outputs. Nevertheless, there are too many ways we do not know and might never know. At the moment, it seems obtaining approximate estimates is the best we can do. There will be situations when we will have to accept that we can not discard several competing theories. Our cognitive shortcomings suggest we should not get attached to any of them, if possible, in order to keep our brains in their best ``truth''-seeking mode.

\bibliography{biblio}     

\end{document}